\documentclass[twocolumn,showpacs,preprintnumbers,amsmath,amssymb]{revtex4}

\usepackage{graphicx}
\usepackage{dcolumn}
\usepackage{bm}

\begin{document}

\title{Effects of thermal- and spin- fluctuations on the band structure of purple bronze Li$_2$Mo$_{12}$O$_{34}$}

\author{T. Jarlborg}
\author{P. Chudzinski}
\author{T. Giamarchi}

\affiliation{
DPMC-MaNEP, University of Geneva, 24 Quai Ernest-Ansermet, CH-1211 Geneva 4,
Switzerland
}

\date{\today}

\begin{abstract}
The band structures of ordered and thermally disordered Li$_2$Mo$_{12}$O$_{34}$ are calculated by use
of ab-initio DFT-LMTO method. The unusual, very 1-dimensional band dispersion obtained in
previous band calculations is confirmed for the ordered structure, and the overall band structure agrees reasonably with existing
photoemission data. Dispersion and bandstructure perpendicular to the main dispersive direction is obtained.
A temperature dependent band broadening is calculated from configurations
with thermal disorder of the atomic positions within the unit cell.
This leads a band broadening of the two bands at the Fermi energy which can become comparable to their energy separation.
The bands are particularly sensitive to in-plane movements of Mo sites far from the Li-sites,
where the density-of-states is highest. The latter fact makes
the effect of Li vacancies on the two bands relatively small. Spin-polarized band results
for the ordered structure show a surprisingly large exchange enhancement on the high DOS Mo sites.
Consequences for spin fluctuations associated with a cell doubling along the conducting
direction are discussed.
\end{abstract}


\maketitle

\section{Introduction.}

The Lithium Purple Bronze, (Li$_{0.9}$Mo$_{6}$O$_{17}$) is
an unusual material which draws attention of research for nearly
three decades \cite{McC,Schlen}. It is a compound
with a rather complicated layered structure \cite{onoda} and
highly 1-dimensional (1D) electronic interactions
\cite{green,whang,popo}. Indeed band calculations with 3D
interactions show an electronic structure with 1D character
\cite{whang,popo}, and with large band dispersion in the direction
perpendicular to the layers. The band structure shows that two
bands are very close together near the crossing of the Fermi energy, $E_F$.

Because of its one dimensional electronic properties, the purple bronze is a
good playground to study the physics of one-dimensional interacting fermionic
systems, which is described by the Tomonaga-Luttinger liquid (TLL) universality class
\cite{QC}. In fact, several experimental results \cite{wang1,wang,hager,xue,gweon}
are consistent with the observation of such one-dimensional physics.

However some deviations from the simple TLL scaling are also
present. Recent results from angular resolved photoemission
spectroscopy (ARPES), done with very high resolution by Wang {\it
et al} \cite{wang1,wang}, indicate some deviations from TLL
predictions. The exponents for $\omega$ and $T$ dependence are
different from predictions from TLL theory, and yet another
exponent is deduced from scanning tunnelling measurements (STM)
\cite{hager}.

It is thus important to examine the possible sources from such
deviations. One possible origin, as with any quasi-one dimensional
system can come from the inter-chain coupling, which drives the
system away from the one dimensional fixed point. We determined in
another paper, starting from the limit of high energy, a
microscopic model incorporating the effects of interactions and
hopping between the chains \cite{cjg}. Such model applicable for
energies above $30$ meV gives results compatible with a Luttinger
liquid description.

In the present paper we focuss on the low energy limit, and
examine using Density Functional Theory (DFT) if additional
ingredients should be incorporated in this previous description
based on a \emph{rigid} band structure for the material. In
particular we examine whether the thermal motion of the atoms in
the solid can lead to a significant enough source of disorder that
could blur the 1D description. Such effects, in three
dimension, are indeed important for the properties of certain classes of compounds
with sharp structures of the density-of-states (DOS) near $E_F$,
such as in the B20 compound FeSi \cite{fesi}, and also for the appearance of the bands
in other materials \cite{wilk,giust,marini}.

We thus reexamine, in this light, in the present paper the band
structure of the purple bronze. In order to ascertain the effects
of the perpendicular hopping we compute carefully the dispersion
in the transverse direction. In addition we examine the effects of
thermal fluctuations and spin fluctuations.

The plan of the paper is as follows. In Sec.~\ref{sec:band} we
describe the method with the results for ordered structure. The
effect of deviations from the ideal atomic structure, due e.g.
from thermal fluctuations, are given in Sec.~\ref{sec:disorder}.
Section Sec.\ref{sec:stat-disor} is dedicated to the effects of
static, substitutional disorder. Results for spin fluctuations and
a discussion of their effects are given in Sec.~\ref{sec:spin}. In
Sec.~\ref{sec:arpes} we present models of how smearing and partial
gaps can modify photoemission intensities.

\section{Band structure.} \label{sec:band}

In this section we apply density functional band theory (DFT) in
order to see to what extent it can explain the unusual
photoemission data. In doing so it is important to note that
effects of thermal disorder and spin-fluctuations may be important
and have to be included in the density functional approach. The
electronic structure of Li$_2$Mo$_{12}$O$_{34}$ (two formula units
of stoichiometric purple bronze) has been calculated using the
Linear Muffin-Tin Method (LMTO, \cite{oka,arb}) in the local
density approximation \cite{lda}(LDA), with special attention to
effects of structural disorder. The lattice dimension and atomic
positions of the structure have been taken from Onoda {\it et al}
\cite{onoda}. The lattice constant in the conducting
$y$-direction, $b_0$, is 5.52 \AA and more than two times larger
along the least conducting $x$-direction $a_0=12.76$ \AA. Thus the
structure consist of well separated slabs. In order to adapt the
LMTO basis for an open structure as purple bronze we inserted 56
empty spheres in the most open parts of the structure. This makes
totally 104 sites within the unit cell. The basis consists of
s-,p- and d-waves for Mo, and s- and p-states for Li, O and empty
spheres, with one $\ell$ higher for the 3-center terms.
Corrections for the overlapping atomic spheres are included. All
atomic
sites are assumed to be fully occupied (except for the case with a
vacant Li atom, see section Sec.\ref{sec:stat-disor}), and they
are all considered as inequivalent in the calculations.
Self-consistency is made using 125 k-points, with more points for
selected paths for the band plots.

\begin{figure}
\includegraphics[height=6.0cm,width=8.0cm]{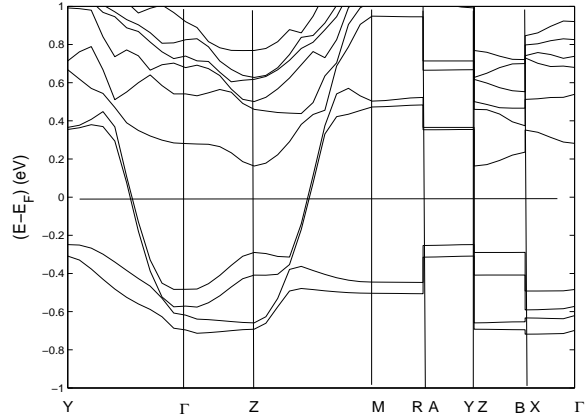}
\caption{Band structure of of purple bronze
Li$_2$Mo$_{12}$O$_{34}$ computed within DFT-LDA approximation for
the rigid structure with parameters taken from Ref.\cite{onoda}.
Bands are shown along symmetry lines in a 1eV window around the
Fermi energy $E_F$.} \label{bndplt}
\end{figure}

\begin{figure}
\includegraphics[height=6.0cm,width=8.0cm]{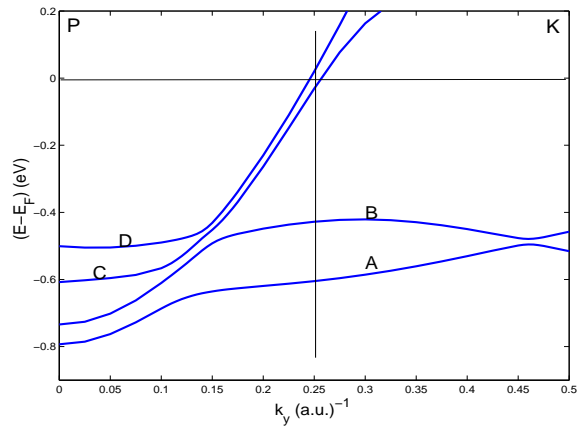}
\caption{Band dispersion along $PK$ (parallel to $\Gamma-Y$
halfway inside the zone) showing that Fermi momentum $k_F$ is very
close to half of the $P-K$ distance (at the vertical line). The
two bands crossing $E-F$ are very close for momenta just below
$k_F$ (for details see Fig.\ref{path5}) but they are separated by
$\sim$ 100 meV for $|k| \leq \frac{1}{2}|k_F|$. 
chosen because the separation between the two bands (C and D) is
the largest for this value of $k_z$. 
Notation as in
Ref.\cite{wang}; C and D corresponds to bands 139 and 140
respectively.} \label{pathPK}
\end{figure}

\begin{figure}
\includegraphics[height=6.0cm,width=8.0cm]{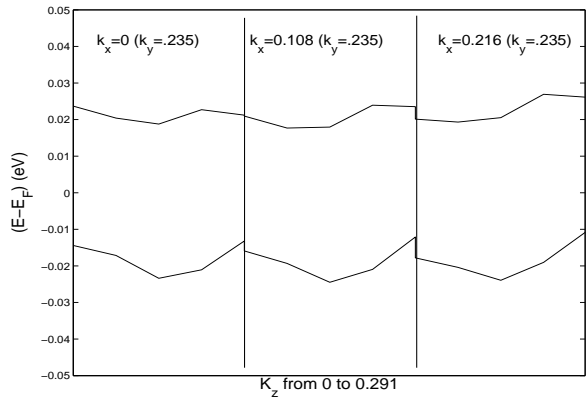}
\caption{Band dispersion of the two bands crossing $E_F$ shown
along the second conducting $z$-direction. Because of lattice
symmetry only a half of a dispersion curve ($k$-values in units 
of 2$\pi$/$b_0$) is shown. Three different cuts of relevant portions of
the Brillouin zone are given.} \label{path5}
\end{figure}

The band structure for the undistorted structure is shown in
Figs.~\ref{bndplt}-\ref{path5}. The total DOS
at $E_F$, $N(E_F)$, is not very large, 2.1 states per
cell and eV. The states at $E_F$ are mainly of Mo-d character
coming from sites far from the region containing Li, where locally
$N(E_F)$ amounts to about 0.2 st./eV/Mo. This is less than 1/3 of
$N(E_F)$ in bcc Mo.

Two bands, number 139 and 140 (counted from the lowest of the
valence bands), are the only ones crossing $E_F$. It was already
pointed out in Ref.\cite{whang} that these two bands originate
from zig-zag chains (oriented along y-axis) which are grouped in
pairs (along z-axis). This is the way each structural slab is
built. The two bands have similar DOS at $E_F$ to within $~$10
percent accuracy. The Fermi velocity, $v_F^y$, along the
conducting $y$-direction is normal as for a good metal, about
5$\cdot 10^5$ m/s, but the ratio between the Fermi velocity in y-
and z-directions, about 50, is compatible with the reported
anisotropic 1D-like resistivity \cite{green}. The velocity along
$\vec{x}$ is even smaller \cite{xyz}.  The bands and
the DOS agree reasonably with the bands calculated by Whangbo and
Canadell \cite{whang} and Popovi\'{c} and Satpathy \cite{popo}
using different DFT methods. In Figs.~\ref{pathPK}-\ref{path5} we
show the details of the bands near $E_F$, where the band
separation and the electronic interactions (or $t$-integrals in a
tight-binding language) can be extracted. The band dispersion
along the conducting the $\Gamma-Y$ (or $P-K$) direction agree
well with the measured results obtained by ARPES \cite{wang} showing a
flattening of the two dispersive bands at about 0.4-0.5 eV below
$E_F$. Other bands are found about 0.25 eV below $E_F$. The
calculated structures A,B,C and D are identified on the bands in
the $P-K$ direction in Fig.~\ref{pathPK} at 0.6 (0.6), 0.4 (0.3),
0.6 (0.5) and 0.5 (0.4) eV below $E_F$, respectively, where the
values within parentheses are taken from the photoemission data of
Ref.~\onlinecite{wang}. Thus, there is an upward shift of the
order 0.1 eV in photoemission compared to the calculated bands.
Such trends are typical, for instance in ARPES on the cuprates,
and can to some extent be attributed to electron-hole interaction
in the excitation process \cite{bost}.  The overall agreement
between different band results and photoemission is reassuring for
this complicated structure, while we now should focus on finer
details of the bands near $E_F$.

The 1-D character of the band structure is revealed from the
comparison between Figs.~\ref{bndplt}-\ref{path5}. While the band
width for the bands crossing $E_F$ is of the order 2 eV in the
$y$-direction, it is not larger than 0.02 eV along $z$ and $x$.
These low values and the shape of the $z$- and $x$-dispersions are
of importance for a Luttinger description of bands close to $E_F$
\cite{cjg}. It should be noted that the in-plane band dispersion
is very small, and that the 2 bands have similar but not identical
shapes along the $z$-direction, see Fig.~\ref{path5}. The
dispersion is quite unusual with minimum in the middle of the
Brillouin zone instead at $q_c =0$. This suggests that there exist
not only one but several equally important, competing
hybridization paths between different chains. A brief analysis of
the structure\cite{xyz} indicates that several paths are possible.
First, for the closest pair of zig-zag chains there is a direct hopping between
Mo(1) and Mo(4) sites, and an indirect one going through Mo(2) and
Mo(5) atoms.
Secondly,  the hopping between these pairs (of chains) always has to go
through Mo(2) atoms, while the next nearest neighbor hopping goes through
M(2) and M(5) atoms.

The direct inter-chain hopping (between Mo(1) and Mo(4) sites) is
weakened because it goes through rather weak $\delta$-bond, while
hoppings through M(2) and M(5) sites are enhanced because their
octahedra are more distorted. The cuts along different $k_b$ are
nearly identical, which shows that this competition does not
depend on the momentum of the propagating wavepacket. The
amplitude of the inter-chain hopping, as computed from band
structure shows that the one dimensionality of the compound is
certain for energies above $\sim 20$ meV, scale of the bare
inter-chain coupling. Below this energy the question is open, and
will be strongly dependent on how the transverse hopping can be
affected by strong correlations \cite{cjg}.

Another check was also made by calculating the bands for the
structural $a$-,$b$-, and $c$-parameters corresponding to low T.
It has been reported that the thermal expansion is quite unusual,
where almost no expansions are found along $b$ and $c$ between 0 K
up to room temperature \cite{santo,luz}. A calculation for the ordered
structure with all structural $a$-parameters downscaled by 0.3
percent, to correspond to the structure at 0 K, was made. The
effects on the bands are very small and not important for the
properties discussed in this work.
The average band separation decreases, but the change is
not significant compared to the original band separation of the
of order 30 meV.

\section{Thermal disorder and zero-point motion.} \label{sec:disorder}

The previous results were obtained by assuming an ideal periodic
structure. Band structure calculations rarely take into
account thermal distortions of the lattice positions. However,
structural disorder due to thermal vibrations is important at high
$T$, and properties for materials with particular fine structures
in the DOS near $E_F$ may even be affected at low $T$
\cite{fesi,ped}. Here for purple bronze, effects of thermal
fluctuations might be pertinent on the degree of dimensionality,
the band overlap between the two bands at $E_F$.

Phonons are excited thermally following the
Bose-Einstein occupation of the phonon density-of-states (DOS),
$F(\omega)$.
The averaged atomic displacement amplitude, $\sigma$, can be calculated
as function of $T$ \cite{zim,grim}.  The result is approximately
that $\sigma_Z^2 \rightarrow 3\hbar\omega_D/2K$ at low $T$ due to zero point motion (ZPM)
and $\sigma_T^2 \rightarrow 3 k_BT/K$ at high $T$
(``thermal excitations''), where $\omega_D$ is a weighted average of $F(\omega)$. The force constant, $K=M_A\omega^2$,
where $M_A$ is an atomic mass (here the mass of Mo is used because of its
dominant role in the DOS), can be calculated as
$K = d^2E/du^2$ ($E$ is the total energy), or it can be taken from experiment.
We use the measurements
of the phonon DOS of the related blue bronze K$_{0.3}$MoO$_3$ \cite{requ} to estimate $K$ and the
average displacements of Mo atoms, as will be explained later.

The individual displacements $u$ follow a Gaussian distribution function.
\begin{equation}
g(u) = (\frac{1}{2\pi\sigma^2})^{3/2} exp(-u^2/2\sigma^2)
\label{Neffeq}
\end{equation}
where $\sigma$, the standard deviation, will be a parameter in the
different sets of calculations. In order to get an estimate to the
effect of such atomic displacements on the band structure, each
atomic site in the unit cell is assigned a random displacement
along $x,y$ and $z$ following the Gaussian distribution function.
Band calculations are made for a total of nine different
disordered configurations. The effects will be a shift of the band
position, to which both ZPM and thermal fluctuations contribute
and a broadening of the bands. Our
calculation, in which we will displace atoms from their natural
position, will also give information on the nature of the band,
and their sensitivity in touching a certain type of atoms.

For these investigations we have performed two types of
simulations of thermal effects: first including all atoms (treated
on equal footing) and second only for atoms around the zig-zag
chain where the DOS at $E_F$ is high. In the latter case, the
distortion, $\sigma_W$, is averaged over those sites only, see
later. These calculations confirm, as expected from the static
band structure, that mostly atoms around the zig-zag chains
contributes to observed effects. No general correlation between
the displacements of nearest neighbors is taken into account, but
extreme values of $u$ are limited in order to avoid that two
atomic spheres make a ``head-on'' collision, which of course would
not occur in the real material. Further refinements of the
disorder could involve different disorder for different atomic
mass, and anisotropic disorder. Purple bronze is a layered
material, 1D-like, and it is probable that vibrational amplitudes
are different perpendicular to the planes compared to within the
layers. However, such information is missing and here we assume
equal isotropic disorder for all atom types. The  present
calculation already gives an estimate of the typical effects of
such atomic displacements.

\begin{figure}
\includegraphics[height=6.0cm,width=8.0cm]{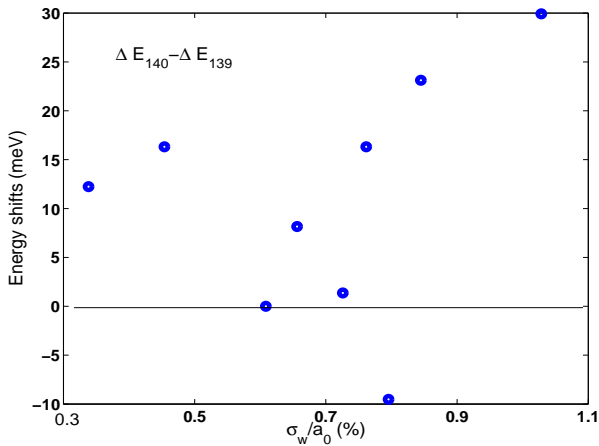}
\caption{The difference of average shifts of the two bands between ordered and disordered structures
as function of the site weighted disorder parameter $\sigma_W$. The zero-point motion corresponds
approximately to $\sigma_w$ lower than  0.7 
means that the two bands move closer for that disorder. Band broadening in the case for $\sigma_W \approx$ 0.7,
when there is no significant average shift, is only because of internal wiggling of the bands, see fig. \ref{fig4}.
}
\label{fig3}
\end{figure}

\begin{figure}
\includegraphics[height=6.0cm,width=8.0cm]{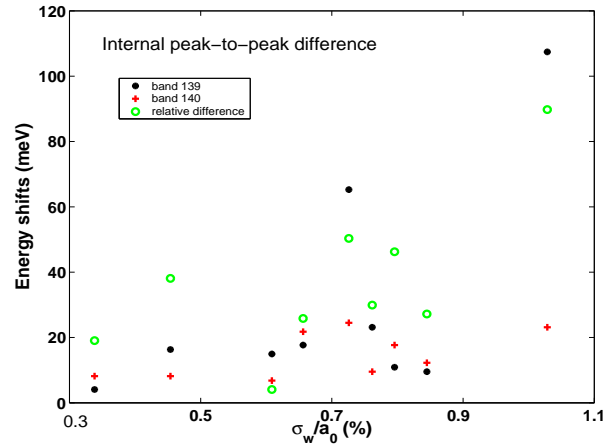}
\caption{Internal peak-to-peak energy difference between undistorted and distorted
bands (black * band 139, red + band 140). The green circles show the maximal difference
between the two previous values. The site weighted disorder parameter $\sigma_W$
is defined in the text.
}
\label{fig4}
\end{figure}

The broadening parameter $\sigma$ depends on $T$ and the
properties of the material. From the experimental data in blue
bronze \cite{requ} we estimate the $T$-dependence of $\sigma$ for
purple bronze. From this we find that $\sigma_{Z}/b_0$ is of the
order 0.7 percent for Mo, and thermal vibrations become larger
than $\sigma_Z$ from about 120 K. For oxygen sites $\sigma_Z/b_0$
is of the order 1 percent. However, as will be discussed later the
band structure is more sensitive to disorder on the Mo sites.

Two measures of the band disorder are shown in
Figs.~\ref{fig3}-\ref{fig4} as function of the weighted
displacements, $\sigma_W$, calculated as the average of
distortions for the 22 sites (far from Li) with the highest
$N(E_F)$. The broadening for each of the two bands is calculated
from the changes in band energies ($\epsilon(k)$) at 25 k-points,
which are found within $\sim$20 meV from $E_F$ in the undistorted
case. The difference between average energy shifts of the two
bands are shown in Fig.~\ref{fig3} as function of $\sigma_W$. This
can be thought as a good measure of the thermally induced effects.
There are more often positive energy differences, indicating that
in average the two bands tend to separate because of the atoms'
random displacements. The net effect is rather weak, we predict
only 10meV difference between room and helium temperature, but it
could be observable. What is more the thermal expansion of the
crystal in b-direction in anomalously weak, so it should not
affect the above result. This outcome is quite unusual, but the
origin of it becomes clear when one analyze the structure of a
crystal. It has been suggested \cite{whang} that $t_{\perp}$
hopping along c-axis is particularly weak in purple bronze because
of certain cancellation in hopping integrals (the $\delta$-bonds
mentioned in Sec.\ref{sec:band}), when the ions reside in a high
symmetry points. Our finding puts that statement on firm ground:
we clearly notice that when ions are slightly shifted the
perpendicular hopping can benefit noticeably.

We can gain even more information when studying the internal
behavior of the two bands separately. Measures of the "wiggling"
of each band because of disorder are displayed in Fig.~\ref{fig4}.
Peak-to-peak differences are given by
$$\Delta E_{max}^j =  |max
[\Delta \epsilon^j(k_n)]-min[\Delta \epsilon^j(k_n)]|$$

$j$ = 139 or 140, where $\Delta \epsilon^j(k_n)$ is the deviation
in energy (with respect to undistorted band) at a certain point
$k_n$ for band $j$. The search for extremes goes through the
set of k-points $k_n$ ($n= [1,25]$). These differences of
extremes are shown for the two bands. For the largest disorder,
band 139 has changed some 100 meV at one k-point relative to the
change of the same band at some other k-point, however this quite
large value has to be taken with caution. In general the bands'
wiggling is of the same order as the average difference between
the two bands (shown on Fig. \ref{fig3}). We suspect that the
origin of wiggling is the same as the origin of increased band
separation, namely the activation of certain overlap integrals.
However, the fact that the wiggling is equally strong as the band splitting implies, at least
within naive tight-binding interpretation, that either the
inter-ladder hopping is equally strong or that the next nearest
neighbor hopping is of the same order as the standard $t_{\perp}$.
This second implication goes along the same lines than the ones
discussed in the context of Fig.~\ref{path5}.

The lower band appears to be more sensitive than the upper band in this respect. In Fig.~\ref{fig4} is
also shown the maximum difference,
$$\Delta\varepsilon =  max|\Delta \epsilon^{139}(k_n)-\Delta \epsilon^{140}(k_n)|$$

It is defined as relative bands deviations at a certain point
$k_n$. In particular case when the two bands would change in the
same way because of thermal disorder (\emph{e.g.} shift
homogenously), then this last value would be equal to the absolute
value of what is shown in Fig.~\ref{fig3}. This is not the case,
which means that the two bands will change quite independently of
each other. In other words, most of the k-points are concerned by
the energy changes, and the wiggling of the two bands is
different. The interpretation of this subtle difference, not
visible within standard DFT band structure result, is quite
difficult. It implies that bands 139 and 140 do not have identical
site dependence. In other words one should not naively assume that
they originate from two zig-zag chains which are very weakly
hybridized as it is commonly done in the literature. The
microscopic complexity of effective interactions, at least at the
lowest energy scales of order $\sim 10meV$, needs to be rich.

In our calculation we have considered displacements of all sites,
but the bands at $E_F$ are mostly sensitive to disorder of Mo
sites with a high DOS. This is reflected by calculations in which
\emph{only} the four Mo with the highest DOS are displaced 0.7
percent of $b_0$ in a few selected directions. When these sites
are displaced perpendicular to the diagonal
($\vec{x},\vec{z}$)-direction the average band shift (equivalent
to the shifts displayed in Fig.~\ref{fig3} for general disorder)
is about 10 meV, while if the displacement is in the directions
parallel to the diagonal the shift is about 50 meV. The effect of
movements in the $\vec{y}$-directions is typically one order of
magnitude smaller. This indicates that the phonon mode whose
displacements are in the plane perpendicular to zig-zag planes
with the movements in the direction of the closest Mo neighbors,
is particularly strongly coupled to electronic liquid.

These values of energy changes can be compared with the
distance between the bands near $E_F$ in the undistorted case, shown
above in Fig.~\ref{path5} for the relevant paths in k-space. The
undistorted bands are separated by 30-40 meV, and the two bands
are confined within 10-15 meV along $k_x$ and $k_z$. The
additional wiggling for distorted cases is of the same order, 20-40
meV,  as seen in Fig.~\ref{fig4}. However, this effect
coexists with a small increase of the band separation. The band separation,
will generally increase when thermal fluctuations are at play.
This is an indication of increased electronic delocalization
within the layers towards a more 3D band structure, but this also
indicates that any disorder will strongly affect carriers
propagation along c-axis. As the temperature increases the thermal
fluctuation effects will increase which means that, if this
picture is valid, the experimentally observed band broadening
should noticeably increase with temperature. This becomes
specially important when thermal activation of phonons becomes dominant at
about 120 K. In addition, the relatively rapid dispersion along
$k_y$ makes the band overlap important in k-space. For instance, a
band separation of 25 meV corresponds to a shift of $k_y$ of only
0.01 of the $\Gamma - Y$ distance.

\section{Static disorder}\label{sec:stat-disor}

In addition to the thermal effects, on which we concentrated in
the previous section, atomic distortions can come from a variety
of other sources. In particular imperfections of the crystal (from
non-stoichiometry, vacancies, site exchange etc.) should also
contribute to the effect. One can also worry about the fact that
one out of ten Li atoms are missing in the real material. The unit
cell considered here contains two Li, and the influence on the
electronic bands from the replacement of one of these with an
empty sphere (without taking structural distortion into account)
is moderately large. The Fermi level goes down because one
electron is removed, but if one neglects the chemical potential
shift he finds an increase of the average band separation between
band 139 and 140 by about 40 meV. This is comparable with largest
calculated effects of thermal disorder, see Fig.~\ref{fig3}.
However, a more realistic estimate is obtained in virtual crystal
calculations where one of the Li is replaced by a virtual atom
with nuclear and electron charges of 2.8 (instead of 3.0 for the
other Li). This set-up has the correct electron count
corresponding to 90 percent occupation of Li. Now the average
shifts of the two bands is much smaller, and the wiggling of the
bands are not even comparable ($\sim$ less than half) to what is
found from ZPM. Therefore it can be expected that effects from
thermal disorder will overcome those from structural disorder in
high quality samples.

The smallness of this effect has important implications if one
looks from the 1D Luttinger liquid perspective. The random Li
vacancies are placed relatively far from zig-zag chains where 1D
liquid resides. This implies that interaction will have Coulomb
character, the small momentum exchange events shall dominate. Thus
substitutional disorder will have primarily forward scattering
character with an amplitude $\approx 15meV$ as determined above.
This situation can be modelled as a Luttinger liquid with forward
disorder. In this case the spectral function is can be given
\cite{cjg}. This implies that substitutional disorder cannot be
invoked to explain phenomena taking place at energy scales larger
than 15 meV. For these larger energies the standard Luttinger
liquid behavior is expected.

\section{Spin fluctuations.} \label{sec:spin}

In Figs.~\ref{bndplt} and \ref{pathPK} it is seen that the two
free-electron like bands cross $E_F$ very close to half of the
$\Gamma-Y$- or $P-K$-distance along the conducting
$\vec{y}$-direction of the structure. A doubling of the real space
periodicity in this direction would open a gap in the DOS near
$E_F$ and lead to a gain in total energy
\cite{htc,Peierls,beni,pytte}, suggesting that this material might
have intrinsic spin density waves type instabilities. In strongly
correlated materials solving these question is of course
complicated, specially in a low dimensional material, since the
proper spin exchanges and quantum fluctuations have to be taken
into account. Some of these issues can be addressed at the level
of the microscopic model derived in Ref.~\onlinecite{cjg}. Here we
look at the possibility of such a spin instability, at the band
structure level and low temperatures, where the two- and three-
dimensional aspects of the system can a priori play a more
important role, and thus the effect of interactions can be reduced
\cite{QC}

A complete verification of how a cell doubling with phonons or
magnetic waves affects the band near $E_F$ would require more
complex band calculations, and will be a demanding undertaking.
Instead we propose at this stage to extract information from
calculations for the 104-site cell, and to apply a free electron
model of the band dispersion in the $y$-direction to see if at all
magnetic fluctuations might be of interest for a band gap. We
noted that displacements of some Mo with the highest local
$N(E_F)$ contribute much to the band distortion. But the size of
the potential shifts at these sites are limited by rather
conservative force constants. Therefore, even if a selected phonon
can contribute to a gap near $E_F$, it might be more effective to
open gaps through spin waves since the potential shifts in this
case can diverge near a magnetic instability. In order to estimate
the strength of the exchange enhancement on Mo we extend our
calculations for the ordered structure to be spin polarized with
an anti-ferromagnetic spin arrangement of the moments on the Mo
with the highest $N(E_F)$. This is made by application of positive
of negative magnetic fields within the atomic spheres of two
groups of four Mo in the cell with the highest DOS.

The propensity for fluctuations of anti-ferromagnetic moments
within the cell is surprisingly large according to the
calculations. The local exchange enhancement on Mo, corresponding
to the Stoner factor for ferromagnetism (i.e. the ratio between
exchange splitting and applied magnetic field), is close to 4.5 in
calculations at low field and temperature. The total energy $E(m)$
is fitted to a harmonic expansion of the moment amplitude $m$,
$E_m = K_m m^2$, where $K_m$ is the ``force constant''. In analogy
with phonon displacement amplitudes one can estimate $m^2 =
k_BT/K_m$ as a measure of the amplitude of moment fluctuations
\cite{htc2}. While phonon distortions $u$ always increase
with $T$, in a Fermi liquid framework one expects that magnetic
moments $m$ will be quenched at a certain $T$ \cite{htc}. This is
because of a self-supporting process where $m$ is a function of
the exchange splitting $\xi$. The latter depends on $m$ and the
local spin density, and the mixing of states above and below $E_F$
given by the Fermi-Dirac function will reduce $m$ at high $T$, so
$m$ and $\xi$ can drop quite suddenly. Calculations at an
electronic temperature of 200K make $K_m \approx 2 eV/\mu_B^2$.
>From this one can estimate that $m$ will be of the order 0.1
$\mu_B$ per Mo at room temperature corresponds to $\xi \approx 0.2
eV$.
Another complication is that the thermal disorder of the lattice
are mixing states across $E_F$ too, and this is another
diminishing factor for spin polarization. Nevertheless, it
suggests that spin fluctuations can play a role at low or
intermediate temperatures. It is interesting to note that the
value of $\xi$ is of the same order of magnitude than the
superexchange parameter calculated from the strong correlations
perspective \cite{Nishi,cjg}. If one compares this value with the
previously determined strength of disorder and effects of thermal
fluctuations, one would realize that spin fluctuations play a much
more prominent role in the low energy physics. Such fluctuations
can potentially lead to pseudogap features in the band dispersion,
thus understanding better their properties is an interesting
challenge, clearly going way beyond the scope of this paper. On
the experimental side, there was only one report of such large gap
($\Delta >10meV$) in the low energy spectrum \cite{xue} and in the
light of more recent experiments \cite{gweon} done with the same
method this finding is highly controversial. An extremely small
gap has been observed in transport experiments \cite{Hussey}, and
probably in STM \cite{hager}, however the spin sector probed by
static susceptibility \cite{green,choi} and muon spectroscopy
\cite{Mandrus} certainly is not gapped. Further theoretical studies are
necessary to understand this situation, where high magnetic
propensity does not lead to a gap for spin excitations.

\section{Comparison with ARPES} \label{sec:arpes}

We show in this section some of the consequences of the band
structure calculated above for the ARPES data. Note that the above
calculation do not take into account the effects of strong
correlations. Thus the deviations from the picture presented below
should thus be direct measure and consequence of such effects.
This will of course depend on the range of temperature and/or
energy.

In order to simulate ARPES intensities for the free-electron bands near $E_F$ we ignore
matrix elements and energy relaxations due to electron-hole interaction. We consider
one-particle excitations directly from the band occupied according to the Fermi-Dirac
distribution, and with pyramidal broadening functions for energy (experimental and intrinsic broadening
due to disorder) and momentum, $k$. The experimental broadenings have the same FWHM values
as in the work of Wang {\it et al}, and we chose to show cuts in momentum of the
same step size as in their work \cite{wang}. The free electron band is fitted to the LMTO
results so that $E_F$ is 0.6 eV at $k_F$. The splitting into
two bands of $\sim$30 meV,
is assumed constant everywhere, which is approximately true for the real bands near $E_F$.
A linear background is added to the intensities in order to make a more
realistic display of photoemission with secondary excitations. Five cuts in momentum below $k_F$,
in equal steps as in Ref.~\onlinecite{wang},
descend to about -0.2 eV. The five cuts appear almost equally spaced in energy, since the
the dispersion is almost linear within the narrow energy interval.

\begin{figure}
\includegraphics[height=6.0cm,width=8.0cm]{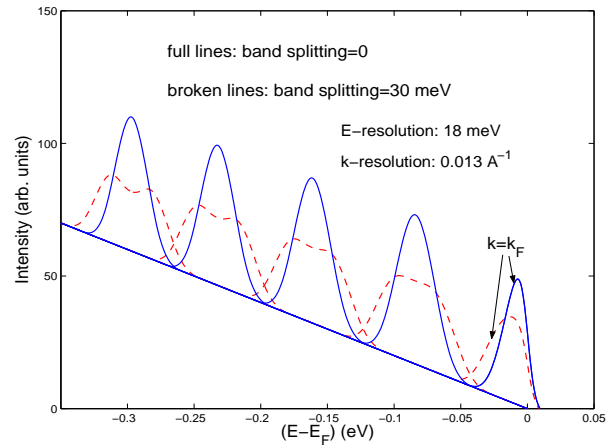}
\caption{Free electron intensities for 5 $k$-values going from $k_F$ in steps of 3.6 percent
of $\Gamma-Y$ added to an arbitrarily chosen background.
The energy and momentum resolutions $\Delta E$ and $\Delta k$ are similar to the experimental values in
ref. \cite{wang}. It is seen that a band splitting of more than $\sim$30 meV should be seen in
photoemission.
}
\label{inten1}
\end{figure}

\begin{figure}
\includegraphics[height=6.0cm,width=8.0cm]{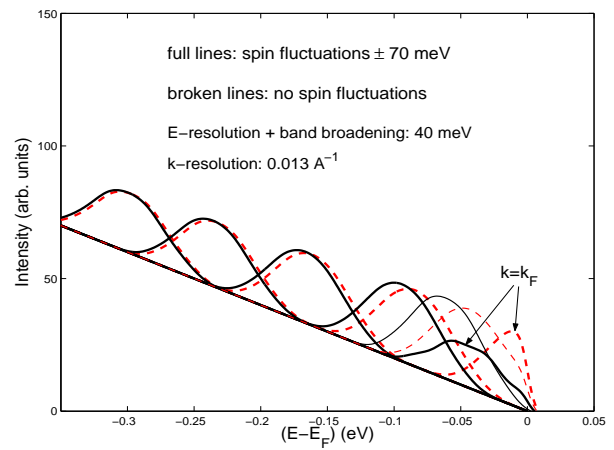}
\caption{Calculated intensities for a band splitting of 30 meV as in fig. \ref{inten1}, but
where an band broadening from ZPM of $\sim$25 meV has been included (broken lines). One intermediate
cut with $k$ at 1.8 percent from $k_F$ is added in this figure. The band broadening
makes the band splitting undetected. Full lines: Including spin fluctuations as described in the text. Note
the energy lowering close to $E_F$.
}
\label{inten2}
\end{figure}

First, as shown in Fig.~\ref{inten1}
the band splitting of about 30 meV should be visible in the ARPES data if only the experimental
broadening functions were at work. Secondly, the broken lines in Fig.~\ref{inten2} show
wider distributions, because of additional broadening coming from the thermal disorder
of the lattice. These distributions are not as wide as in Ref.~\cite{wang}. This strongly
suggests that other physical effects (effects of interactions, static disorder due to non-stoichiometry, spin fluctuations)
are at play.

As an extreme case we can consider a cell doubling (e.g. due to
spin ordering) which would create a gap near  $k_y = k_F$. The
full lines in Fig.~\ref{inten2} show what would happen to the
spectra if $\xi$ in the free-electron model goes up to about 70
meV.  In this case it is seen that the band retracts from $E_F$,
since the peaks for $k$ closest to $k_F$ are lower in energy than
the broken lines.  Far below $E_F$ there is not much difference
between full and broken lines. In the end the presence of a gap
leads to a more non-uniform energy distribution of the five k-cuts
of the intensity. Comparison of the experimental features, with
the one obtained by such a band structure analysis could thus be
useful to investigate the low energy properties of purple bronze,
and in particular the existence of a gap or pseudogap in the
dispersion relation. The analysis in Ref.~\cite{wang} revealed a
$T^{0.6}$ scaling of the intensity near $E_F$ over a wide
$T$-interval. Ascertaining whether this behavior at low energy
comes from one dimensional fluctuations or some pseudogap regime
is an important question.

\section{Conclusion.}

In this paper we have reexamined the band structure of purple
bronze in the ordered lattice and found that the main features of
the bands agree well with previous calculations and ARPES results.
In particular, only two bands cross the Fermi level. In our study
we focus on these two bands. The perpendicular interactions along
$z$ and $x$ (``perpendicular hoppings'') are weak and appear to
lay out the conditions of TLL-like behavior of the bands near
$E_F$, for energy down to at least $20$ meV.

We have investigated whether additional effects, giving small
energy distortions at these energy scales could cause an overlap
between these two bands at low $T$. The energy separation between
the two essentially one dimensional bands increases below $E_F$,
as can be seen in Fig.~\ref{bndplt} along the $\Gamma - Y$
direction. For instance, the two considered bands are separated
$\sim 0.1$ eV at 0.4 eV below $E_F$, similarly to what is seen in
ARPES at this energy \cite{wang}. When bands are closer to $E_F$
then they are also closer together, however as shown in
Sec.\ref{sec:arpes} within single particle picture (LDA-DFT) they
should be distinguishable at low $T$ provided that intrinsic
disorder is not too large and if the matrix elements for
transitions from both bands are comparable \cite{matr}. As $T$
increases beyond $\sim$120 K or more, the band broadening
increases because of additional thermal disorder so that the two
bands seem to merge. We find that the thermal effects can
potentially play a role in the band separations however one has to
keep in mind the temperature dependence of the effect: the bands
becomes less and less distinguishable as the temperature
increases. There is obviously a complicated cooperation between
increase in splitting and wiggling but in any case the following
effect can be used to verify experimentally this picture: as the
temperature increases the ARPES lines should become noticeably
thicker.

Additional calculations for Li deficient purple bronze show no
important modifications of the bands crossing $E_F$, at least as
long as it is not associated with static disorder. Likewise the
band structure for a unit cell with modified $x/y$- and
$x/z$-ratios, appropriate for the structure at room temperature,
is not very different from what is shown in Fig.~\ref{bndplt}.

Recent photoemission result of Wang {\it et al} \cite{wang}, made
at low $T$, only detects one band at $E_F$. The interpretation of
these result is still an exciting issue given the fact that
electron-electron interactions can also suppress tunnelling and
reinforce the one dimensionality of the material \cite{cjg}. What
we showed in this paper is that at a one body level, the effect of
distortion of the electronic structure, also go in the direction
of a smearing of the difference between the two bands and
thickening observed ARPES spectra. Disentangling the two effects
is thus an important question, and can potentially be done by
comparing the predicted band separations, at the level of band
structure, with the actual ARPES data.

We have also explored the possibility of anti-ferromagnetic spin
fluctuations. Here, our investigations are not complete, but our
first results show surprisingly large anti-ferromagnetic exchange
enhancements on Mo within the basic unit cell. This and other
facts motivate further studies of fluctuations within larger
cells, in order to see if they can lead to gap or pseudogap
features near $E_F$.

It is interesting to note that the existence of thermal
fluctuations and imperfections of the atomic structure, makes --
at the band structure level -- the system more three dimensional.
This effect will be in competitions with the renormalization of
the inter-chain hopping coming from the electron-electron
correlations. The competition between these two effects leads to
an intermediate and low temperature physics which is still
mysterious in this compound.

Aknowlegments: We would like to thank Jim Allen and Enric Canadell for shearing their knowledge
about purple bronze with us. We also aknowlege an additional insight provided by Sachi Satpathy, Tanusri Saha-
Dasgupta and Maurits W. Haverkort. This work was supported by the Swiss NSF under
MaNEP and Division II.


\begin{thebibliography}{10}
\bibitem{xyz} We adopt the same conventions as in refs. \cite{onoda,popo} for the structure, site-notation
and k-space, so that $\vec{x}$ and $\vec{z}$ are parallel and $\vec{y}$ perpendicular to the planes.
\bibitem{matr} In ref. \cite{wang} the intensity of the upper band is much lower than of the lower one.
\end{thebibliography}
\end{document}